\documentclass[twocolumn,amsmath,amssymb,aps,pra,superscriptaddress]{revtex4}

\usepackage{array}
\usepackage{amsmath,amssymb}
\usepackage{dcolumn}
\usepackage{bm}
\usepackage{color}
\usepackage{xcolor}
\usepackage{ulem}
\usepackage[utf8]{inputenc}
\usepackage{hyperref}
\usepackage{comment}
\usepackage{mathtools}
\usepackage{natbib}
\usepackage{url}
\usepackage{times}
\usepackage{graphicx}
\usepackage{subfigure}
\usepackage{enumitem}
\usepackage{epstopdf}
\usepackage{xcolor}
\usepackage{ulem}
\usepackage{footnote}
\usepackage{comment}
\usepackage{dsfont}

\newcommand{\myifelse}[3]{\def\FirstArg{#1}\ifx\FirstArg\empty#2\else#3\fi}
\newcommand{\createpsi}[1][]{%
	\def\FirstArg{#1}
	\ifx\FirstArg\empty
	{\hat{\Psi}^\dagger}
	\else
	{\hat{\Psi}_{#1}^\dagger}
	\fi
}

\newcommand{\kron}[2][]{%
	\def\FirstArg{#1}
	\ifx\FirstArg\empty
	\delta_{#2}
	\else
	\delta^{(#1)}(#2)
	\fi
}

\DeclarePairedDelimiter\ket{\lvert}{\rangle}
\DeclarePairedDelimiterX\braket[2]{\langle}{\rangle}{#1 \delimsize\vert #2}

\hypersetup{
    colorlinks=true,
    linkcolor=blue,
    filecolor=magenta,      
    urlcolor=purple,
    citecolor=cyan
}

\usepackage{graphicx}
\usepackage{subfigure}
\usepackage{mathrsfs} 
\usepackage{bm}
\usepackage{mathtools}
\usepackage{epstopdf}
\begin{document}

\title{Anomalous buoyancy of quantum bubbles in immiscible Bose mixtures} 
\author{Daniel Edler}
\affiliation{Institut f\"{u}r Theoretische Physik, Leibniz Universit\"{a}t, 30167 Hannover, Germany}
\author{Luis A.~Pe{\~n}a~Ardila}\thanks{luis.ardila@itp.uni-hannover.de}
\affiliation{Institut f\"{u}r Theoretische Physik, Leibniz Universit\"{a}t, 30167 Hannover, Germany}
\author{Cesar R. Cabrera}
\affiliation{Fakultät für Physik, Ludwig-Maximilians-Universität, Schellingstr. 4, D-80799 München, Germany}
\affiliation{Munich Center for Quantum Science and Technology (MCQST), Schellingstr. 4, D-80799 München, Germany}
\affiliation{Max-Planck-Institut für Quantenoptik, Hans-Kopfermann-Strasse 1, D-85748 Garching, Germany}
\author{Luis Santos}
\affiliation{Institut f\"{u}r Theoretische Physik, Leibniz Universit\"{a}t, 30167 Hannover, Germany}

\begin{abstract}
Buoyancy is a well-known effect in immiscible binary Bose-Einstein condensates. Depending on the differential confinement experienced by the two components, a bubble of one component 
sitting at the center of the other eventually floats to the surface, around which it spreads either totally or partially. We discuss how quantum fluctuations may significantly change the volume and position of immiscible bubbles. We consider the particular case of two miscible components, forming a pseudo-scalar bubble condensate with enhanced quantum fluctuations (quantum bubble), immersed in a bath provided by a third component, with which they are immiscible. We show that in such a peculiar effective binary mixture, 
quantum fluctuations change the equilibrium of pressures that define the bubble volume and modify as well the criterion for buoyancy. Once buoyancy sets in, in contrast to the mean-field case, quantum fluctuations may place the bubble at an intermediate position between the center and the surface. At the surface, the quantum bubble may transition into a floating self-bound droplet.
\end{abstract}

\maketitle




\section{Introduction}
\label{sec:intro}

Quantum fluctuations typically play a negligible role in weakly-interacting Bose gases. This is because the dominant beyond-mean-field correction resulting from 
the zero-point motion of the Bogoliubov excitations, the so-called Lee-Huang-Yang~(LHY) correction~\cite{Lee1957}, is much smaller than the mean-field energy. As pointed out in Ref.~\cite{Petrov2015}, the situation is radically different in Bose-Bose mixtures, in which the mean-field term and the LHY correction depend in a different way with respect to the inter- and intra-component interactions. Remarkably, repulsive intra-component interactions may compensate attractive inter-component ones to quasi-cancel the mean-field energy, hence 
enhancing the role of quantum fluctuations.  Under those conditions, quantum fluctuations, as a result of their repulsive character, may prevent collapse due to the 
steeper density scaling of the LHY term compared to the mean-field one. As a result, the mixture forms an ultra-dilute self-bound liquid, which has received the name of quantum droplet. These droplets have been experimentally realized in both homonuclear~\cite{Cabrera2018, Semeghini2018} and heteronuclear~\cite{DErrico2019} mixtures.
Quantum droplets have been also observed in dipolar gases~\cite{Review2021,Review2022}, in which the competition of short-range and dipolar interactions quasi-cancels the mean-field energy. Quantum stabilization hence results as well in the formation of dipolar quantum droplets~\cite{Kadau2016,Chomaz2016,Schmitt2016}. Interestingly, due to the anisotropy and non-locality of the dipolar interaction, confinement leads  to the formation of a droplet array~\cite{Wenzel2017}, which for properly fine-tuned contact interactions forms a supersolid~\cite{Tanzi2019, Boettcher2019, Chomaz2019}. 

The miscibility of a binary Bose mixture depends on the nature of the inter- and intra-component interactions. 
For a homogeneous system, miscibility is given by the ratio $\Gamma = g_{12}/\sqrt{g_{11}g_{22}}$, between the inter-component coupling constant $g_{12}$ and the intra-component ones $g_{11}$ and $g_{22}$. If $\Gamma>1$ the system enters the immiscible regime characterized by phase separation. The presence of external confinement significantly affects the 
miscibiliity and spatial distribution of binary mixtures~\cite{Ho1996,Pu1998,Timmermans1998,Hall1998,Ohberg1999}. In particular, for an immiscible mixture, depending on the 
relation between the confinement harmonic frequencies $\omega_{1,2}$ of the two components, a bubble of component 1 may sink to the center of the trap or float rather to the surface of  component 2,  in a process that resembles buoyancy in ordinary fluids. Similar to Archimedes' principle, buoyancy in a binary Bose mixture is controlled by the equilibration of the pressures inside and outside the bubble. Buoyancy sets in, approximately, when $\omega_1/\omega_2<(g_{11}/g_{22})^{1/4}$~\cite{Timmermans1998}.   

In this paper, we are interested in the properties of an immiscible binary mixture when one of the two components has enhanced quantum fluctuations.
We consider specifically an experimentally feasible scenario of a three-component Bose gas. Components 1 and 2 are miscible and in the regime of mean-field quasi-cancellation, 
and hence they behave as an effective single-component, which we call the 1-2 gas, in which quantum fluctuations play a relevant or even dominant role. Component 3 is immiscible with respect to the 1-2 gas. We are particularly interested in how quantum fluctuations modify the properties of a 1-2 bubble~(quantum bubble) in a bath given by component 3. 
We show that quantum fluctuations change the bubble pressure, changing the relation between bubble volume and bath density, and the buoyancy criterion. 
Moreover, in contrast to the mean-field case, quantum fluctuations may arrest buoyancy, placing the quantum bubble at an intermediate position between the center and the bath surface. 
In addition, at the surface, the quantum bubble may transition into a floating self-bound droplet. Our results show that, in addition to allowing for quantum droplets, quantum fluctuations 
may significantly affect other well-established properties of quantum mixtures, a result which could be relevant as well for immiscible dipolar mixtures~\cite{Bland2022}. 

The structure of the paper is as follows. In Sec.~\ref{Sec:Model} we introduce the three-component mixture, including the effect of quantum fluctuations.
Section~\ref{Sec:Pressures} discusses how quantum fluctuations affect the equilibrium of pressures that maintains the quantum bubble. 
Section~\ref{Sec:Volume} analyzes the dependence of the bubble volume with respect to the bath density resulting from the modified equilibrium conditions.
The anomalous buoyancy of quantum bubbles is discussed in Sec.~\ref{Sec:Buoyancy}. In Sec.~\ref{Sec:Experiments} we discuss a possible experimental realization.
Finally, in Sec.~\ref{Sec:Conclusions} we summarize our conclusions.




\section{Model}
\label{Sec:Model}

\subsection{Hamiltonian and elementary excitations}

Interactions are determined by the $s$-wave scattering lengths $a_{\sigma\sigma'}$, with $\sigma, \sigma'=1,2,3$. Motivated by experimental considerations~(see Sec. VI), we focus on the case in which the three components have equal mass $m$. The interactions are hence characterized by the coupling constants $g_{\sigma\sigma'}=4\pi\hbar^2 a_{\sigma\sigma'}/m$. The coupling constants $g_{11}$, $g_{22}$ and $g_{12}$ are such that components 1 and 2 are miscible 
and in the regime of mean-field quasi-cancellation~\cite{Petrov2015}. The bath is characterized by the coupling $g_{33}$. The inter-component coupling constants $g_{13}$ and $g_{23}$ are supposed to be large and repulsive, ensuring immiscibility between component 3 and the 1-2 gas. 

In absence of confinement, the system is determined by the Hamiltonian:
\begin{eqnarray}
\mathcal{H}&=&\sum_{\lambda=1}^{3}\int d^3r \, \hat{\psi}_{\lambda}^{\dagger}(\mathbf{r}) \, \left ( \frac{-\hbar^2\nabla^2}{2m}\right ) \,  \hat{\psi}_{\lambda}(\mathbf{r}) \nonumber \\
&+&\sum_{\lambda,\lambda'}\frac{g_{\lambda\lambda'}}{2}\int d^3 r \, \hat{\psi}_{\lambda}^{\dagger}(\mathbf{r})\hat{\psi}_{\lambda'}^{\dagger}(\mathbf{r})\hat{\psi}_{\lambda'}(\mathbf{r})\hat{\psi}_{\lambda}(\mathbf{r}).
\label{eq:Hint}
\end{eqnarray}
We perform the Fourier transform $\hat{\psi}_{\lambda}(\mathbf{r})=\frac{1}{\sqrt{V}}\sum_{\mathbf{k}}e^{-i\mathbf{k\cdot r}}\hat{a}_{\mathbf{\lambda,k}}$, with $V$ the quantization volume, 
and introduce the Bogoliubov transformation 
\begin{equation}
\hat{\beta}_{\mathbf{\alpha k}}=\sum_{\lambda}\left[u_{\alpha\lambda}(\mathbf{k})\hat{a}_{\lambda,\mathbf{k}}
-v_{\alpha\lambda}(\mathbf{k})\hat{a}_{\lambda,\mathbf{-k}}^{\dagger}\right],
\end{equation}
where the coefficients $u_{\alpha\lambda}(\bf k)$ and $v_{\alpha\lambda}(\bf k)$, and the corresponding eigenenergies $\xi_\alpha$ result from the solution of the Bogoliubov-de Gennes equations,  $\xi_{\alpha}(k)\hat{\beta}_{\mathbf{\alpha k}}=\left[\hat{\beta}_{\mathbf{\alpha k}}, \mathcal{H}\right]$. Employing  
$f^\pm_{\alpha\lambda}(\mathbf k) = u_{\alpha\lambda}(\bf k) \pm v_{\alpha\lambda}(\bf k)$, we may express these equations in the form:
\begin{eqnarray}
\label{eq:tbm_bdg}
\xi_\alpha f^-_{\alpha\lambda}(\mathbf k) &=& \sum_{\lambda'} \left [ \delta_{\lambda\lambda'} \epsilon(k)+ 2g_{\lambda\lambda'} \sqrt{n_\lambda n_{\lambda'}}\right ] f^+_{\alpha\lambda'}(\mathbf k), \\
\xi_\alpha f^+_{\alpha\lambda}(\mathbf k) &=&\epsilon(k) f^-_{\alpha\lambda}(\mathbf k),
\end{eqnarray}
where $\epsilon(k) = \hbar^2k^2/2m$. Combining these equations, we obtain
\begin{equation}
\xi_\alpha^2 \mathbf{f}_\alpha (\mathbf{k} ) = \epsilon(k) \left [ \epsilon(k) \hat{\mathds{1}} + 2 g_{11} n_1 \hat{\mathbf{U}}(P_{2},P_{3}) \right] \mathbf{f}_\alpha (\mathbf{k} ),
\end{equation}
where $\left ( \mathbf{f}_\alpha (\mathbf{k} ) \right )_\lambda = f^-_{\alpha\lambda}(\mathbf{k} )$, $P_{j=2,3}={N_{j}/N_{1}}$, and 
\begin{equation}
\hat{\mathbf{U}}(P_{2},P_{3}) = \begin{pmatrix}
1 & \frac{g_{12}}{g_{11}}\sqrt{P_{2}} & \frac{g_{13}}{g_{11}}\sqrt{P_{3}}\\
\frac{g_{12}}{g_{11}}\sqrt{P_{2}} & \frac{g_{22}}{g_{11}}P_{2} & \frac{g_{23}}{g_{11}}\sqrt{P_{2}P_{3}}\\
\frac{g_{13}}{g_{11}}\sqrt{P_{3}} & \frac{g_{23}}{g_{11}}\sqrt{P_{2}P_{3}} & \frac{g_{33}}{g_{11}}P_{3}
\end{pmatrix}.
\label{eq:MatrixU}
\end{equation}
We may then express the Bogoliubov energies in the form:
\begin{equation}
\xi_{\alpha}^{2}(k)=\epsilon(k)\left[\epsilon(k)+2g_{11}n_{1}F_{\alpha}(P_{2},P_{3})\right],
\label{eq:Spectrum}
\end{equation}
with $F_\alpha$ the eigenvalues of $\hat{\mathbf{U}}(P_{2},P_{3})$.

\subsection{Quantum fluctuations}

At the mean-field level, the chemical potential of component $\lambda$ is given by $\mu_{\lambda}^{MF}=\sum_{\lambda'}g_{\lambda\lambda'}n_{\lambda'}$. 
In the following, we will obtain the correction to this chemical potential induced by quantum fluctuations, i.e. the generalization of the well-known LHY 
correction~\cite{Lee1957} to the three-component case under consideration. In order to do so, we employ the formalism introduced by Hugenholz and Pines~\cite{Hugenholz1959}, 
which avoids in a natural way the ultraviolet divergence that results from the usual Bogoliubov treatment, and which is typically cured by considering the 
second-Born approximation of the coupling constants. This formalism, which is based on a Green's function formalism, has been recently applied for the treatment of low-dimensional gases~\cite{Edler2017,Igl2018}, and dipolar mixtures~\cite{Bisset2021}.  
Within the Hugenholz-Pines formalism, the correction $\epsilon_\text{LHY}$  of the energy density of the ground-state of the Bose mixture may be evaluated from 
the knowledge of the Bogoliubov spectrum by means of the differential equation: 
\begin{equation}
\epsilon_{\text{LHY}}-\frac{1}{2}\sum_{\lambda=1}^{3}n_{\lambda}\frac{\partial\epsilon_{\text{LHY}}}{\partial n_{\lambda}}=\chi
\label{eq:HP}
\end{equation}
where
\begin{equation}
    \chi=-\frac{1}{2}\int\frac{d^{3}k}{(2\pi)^{3}}\sum_{\alpha}\frac{[\xi_{\alpha}(k)-\epsilon(k)]^{3}}{4\xi_{\alpha}(k)\epsilon(k)}.
\label{eq:xi}
\end{equation}
Plugging  Eq.~\eqref{eq:Spectrum} into Eq.~\eqref{eq:xi} we obtain that 
\begin{equation}
\chi=-\frac{64}{15}\sqrt{\pi}\frac{\hbar^{2}}{m}(n_{1}a_{11})^{5/2}\sum_{\alpha}F_{\alpha}^{5/2},
\end{equation}
and substituting into Eq.~\eqref{eq:HP} we obtain: 
\begin{equation}
\epsilon_\text{LHY}=\frac{256}{15}\frac{\hbar^{2}}{m}\sqrt{\pi}(n_{1}a_{11})^{5/2}\sum_{\alpha}F_{\alpha}^{5/2}. 
\end{equation}
Hence the beyond-mean field correction to the chemical potential of component $\lambda$ reads:
\begin{equation}
\mu_{\lambda}^{\text{LHY}}=\frac{\partial}{\partial n_{\lambda}}\epsilon_\text{LHY}=\frac{32}{3\sqrt{\pi}}g_{11}\left(n_{1}a_{11}\right)^{3/2}Q_{\lambda},
\label{eq:mu}
\end{equation}
with
\begin{eqnarray}
Q_{1}&=&\sum_{\alpha}F_{\alpha}^{3/2}\left(F_{\alpha}-P_{2}  \frac{\partial}{\partial P_{2}}F_{\alpha} P_{3}  \frac{\partial}{\partial P_{3}}F_{\alpha}  \right), \\
Q_{2}&=&\sum_{\alpha}F_{\alpha}^{3/2} \frac{\partial}{\partial P_{2}}F_{\alpha}, \\
Q_{3}&=&\sum_{\alpha}F_{\alpha}^{3/2} \frac{\partial}{\partial P_{3}}F_{\alpha}.
\end{eqnarray}
The matrix $\mathbf{U}$ in ~Eq.~\eqref{eq:MatrixU} determines the mean-field stability of the mixture.  If all its eigenvalues are positive, the mixture is fully miscible. 
For the purposes of this work, we restrict ourselves to the case where component 3 and components 1-2  are immiscible. Hence 
in the 1-2 region, we may assume $P_3=0$, and we recover the known LHY corrections for the 1-2 mixture~\cite{Petrov2015}. Quantum corrections 
play a crucial role for the 1-2 mixture if the mean-field interactions quasi-cancel, i.e. if $\delta a \equiv a_{12} + \sqrt{a_{11} a_{22}} \simeq 0$. 
If $\delta a <0$, quantum fluctuations are crucial since they stabilize the 1-2 gas against collapse~\cite{Petrov2015}, but even for $\delta a\geq 0$ they play an important role, 
as discussed below.

\subsection{Coupled extended Gross-Pitaevskii equations}

In order to investigate the properties of a spatially inhomogeneous mixture, we use the results obtained for an homogeneous mixture, and 
employ local-density approximation arguments for treating the LHY correction~\cite{Petrov2015}. The use of the local-density approximation is justified by the fact that 
the LHY correction is mostly contributed by excitations with wavelengths shorter than the typical length of density variation. 
We obtain in this way a set of three coupled extended Gross-Pitaevskii equations~(eGPEs):
\begin{eqnarray}
\tilde\mu_\lambda \psi_{\sigma}(\vec{r})&=&\Big[-\frac{\hbar^{2}\nabla^{2}}{2m}+\sum_{\lambda'}g_{\sigma\sigma'}n_{\sigma'}(\vec{r}) \nonumber \\
&+& \mu_{\sigma}^{\text{LHY}} \left [ n_{\sigma'} (\vec r) \right ] \Big] \psi_{\sigma}(\vec{r}),
\label{eq:Coupled-eGPE}
\end{eqnarray}
with $n_\sigma(\vec r) = |\psi_{\sigma}(\vec{r})|^{2}$. Similar equations have been employed in binary mixtures~\cite{Petrov2015} and dipolar gases~\cite{Waechtler2016}, providing very good qualitative, and to a large extent quantitative, agreement with experiments.




\section{Equilibrium of pressures}
\label{Sec:Pressures}

We consider at this point a spherical homogeneous 1-2 bubble of volume $V$, with $N=N_{1}+N_{2}$ particles and polarization $P=N_2/N_1$, placed 
in an otherwise homogeneous bath of component 3, with particle density $n_3$. We determine the relation between the bath density $n_3$ 
and the bubble density $n=N/V$, which is established by an equilibrium of pressures. We are particularly interested in how quantum fluctuations modify such an equilibrium. 

The 1-2 contribution to the bubble energy is
\begin{equation}
E_{12}(V)=\frac{1}{2}G(P)\frac{N^{2}}{V}+\gamma(P) g_{11}a_{11}^{3/2}\frac{N^{5/2}}{V^{3/2}},
\label{E12}
\end{equation}
where the first and second terms correspond, respectively, to the mean-field and LHY corrections, and 
\begin{eqnarray}
G(P)&=&\frac{g_{11}+g_{22}P^{2}+2g_{12}P}{(1+P)^{2}}, \\
\gamma(P)&=&\frac{64}{15\sqrt{\pi}}\frac{f\left (\frac{g_{12}^2}{g_{11} g_{22}},\frac{g_{22}}{g_{11}}P \right )}{(1+P)^{5/2}},
\end{eqnarray}
with 
\begin{eqnarray}
f(x,y)= \frac{1}{4\sqrt{2}} \sum_\pm \left( 1 + y \pm \sqrt{(1-y)^2 + 4xy} \right)^{5/2}
\label{eq:f}
\end{eqnarray}
As already mentioned, components 1 and 2 form an effective scalar component, the 1-2 gas, characterized by an effective scattering length $a(P)$, with $G(P)=\frac{4\pi\hbar^2 a(P)}{m}$, and by enhanced quantum fluctuations that dominate the bubble properties in the mean-field energy quasi-cancels, i.e. if $G(P)\simeq 0$.

Due to immiscibility, the bubble induces a hollow spherical cavity of volume $V$ in the bath. 
The change induced by the cavity in the bath energy is 
$\Delta E_3 = E_{C}(N_3)-E_{NC}(N_3)$, with $E_C(N_3)$ the energy of a bath of $N_3$ particles with the hollow cavity, and $E_{NC}(N_3)$ the energy of the bath without 
the cavity. Note that $E_{C}(N_3)\simeq E_{NC}(N_3+\delta N_3)-\xi_3 V$, where $\delta N_3=n_3 V$, and 
$\xi_3 =\frac{1}{2}g_{33}n_3^2$ is the energy density of the 
third component~(where we have neglected LHY corrections, which for the single component in the bath are assumed as negligible compared to the mean-field energy).
In turn, $E_{NC}(N_3+\delta N_3)\simeq E_{NC}(N_3)+\mu_3\delta N_3$, with $\mu_3=g_{33}n_3$ the chemical potential of the bath.
We may hence write: 
\begin{equation}
\Delta E_3 (V)= \frac{1}{2}g_{33} n_3^2 V.
\end{equation}
The energy associated to the bubble is hence $E_{12}(V)+\Delta E_3(V)$. Minimizing it with respect to $V$, we obtain the equation 
for the equilibrium between the inner pressure $P_{12}=-\partial_V E_{12}$ and the outer bath pressure $P_3=-\partial_V \Delta E_3$: 
\begin{equation}
G(P)n^2 + 3\gamma(P) g_{11} a_{11}^{3/2} n^{5/2} = g_{33} n_3^2.
\label{eq:Equilibrium}
\end{equation}




\section{Quantum bubble in an homogeneous bath}
\label{Sec:Volume}

\subsection{Scaling of the bubble volume with the bath density}

Equation~\eqref{eq:Equilibrium} determines the bubble volume for a given bath density. 
In the mean-field regime, in which we may neglect the effect of quantum fluctuations in the bubble, the equilibrium of pressures results in 
the known expression~\cite{Timmermans1998} 
\begin{equation}
n_{\mathrm{MF}}=\sqrt{\frac{g_{33}}{G(P)}}n_3, 
\label{eq:nMF}
\end{equation}
and the bubble volume is inversely proportional to $n_3$. 

The situation changes significantly when the 1-2 mean-field interactions quasi-cancel. 
For $G(P)=0$ and sufficiently large density, the LHY energy dominates 
the bubble energy (we call this the LHY bubble regime), and the equilibrium of pressures leads to an anomalous dependence:
\begin{equation}
n_{\mathrm{LHY}}=\left ( \frac{g_{33}}{3\gamma(P)g_{11}a_{11}^{3/2}} \right )^{2/5} n_3^{4/5}.
\label{eq:nLHY}
 \end{equation}
The volume of a LHY bubble scales thus as $n_3^{-4/5}$.

In contrast, for $G(P)=0$, if the bubble density is too low, we can neglect the effect of quantum fluctuations and the bubble energy is dominated by the single-particle (kinetic) contribution,  
associated to the inhomogeneity of the bubble wavefunction within the bath cavity, which we have up to now neglected. Approximating that the bubble is in a spherical hard-wall 
cavity of volume $V$, the bubble energy is $\frac{\hbar^2\pi^2}{2m}\left ( \frac{4\pi}{3V}\right )^{2/3}$, leading to an inner pressure
$P_{12}=\frac{\pi^2}{3}\left ( \frac{4\pi}{3}\right )^{2/3}\frac{\hbar^2N}{m V^{5/3}}$. Equating $P_{3}=\frac{1}{2}g_{33}n_3^2$ results hence in a scaling $V\propto n_3^{-6/5}$.



\begin{figure}[t!]
\begin{centering}
\includegraphics[width=\columnwidth]{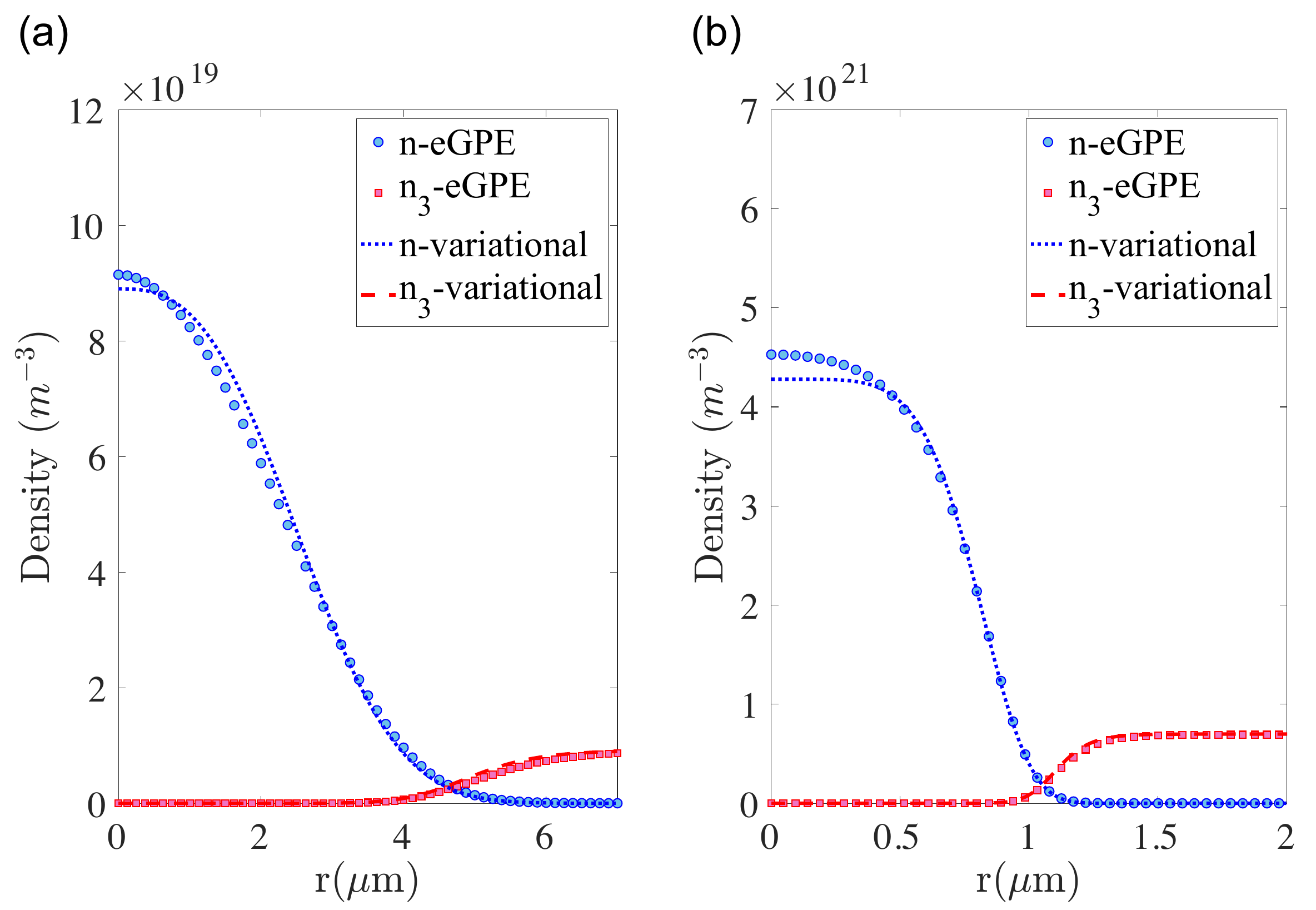}
\caption{Density profile of the 1-2 gas~(circles) and of the bath~(squares) obtained from the coupled eGPEs, for 
$(a_{11},a_{22},a_{13},a_{23},a_{33})=(34.44,82,172,172,60)a_{0}$, $N=10^{4}$, $P=\sqrt{a_{11}/a_{22}}$, for two different bath densities $n_3=9\times10^{18}m^{-3}$~(a) and $ 7\times10^{20}m^{-3}$~(b). 
Dashed lines indicate the corresponding variational results.}
\label{Fig:1}
\end{centering}
\end{figure}


\subsection{Variational formalism}

Quantum fluctuations hence significantly modify the bubble volume, and its scaling with the bath density. 
In order to investigate the bubble properties we employ the coupled eGPEs introduced in Sec.~\ref{Sec:Model}.  
We consider a spherical hard-wall numerical box of radius $R_B$, with $N_3=n_3 \frac{4\pi}{3}R_B^3$ particles in the bath. 
We consider a bubble at the center of the numerical box, with $N=N_1+N_2$ particles. We assume $P=\sqrt{a_{11}/a_{22}}$ here and also in the rest of the paper. 
Although this is a necessary condition for a 1-2 self-bound droplet~\cite{Petrov2015}, 
it is not necessary in our case, in which the bubble is surrounded by a bath. We assume it however for simplicity of the resulting expressions.

In addition, although the eGPE formalism permits a good characterization of the quantum bubble, a simpler variational formalism, discussed in the following, is in very good agreement 
with the eGPE calculation, allows for a quick simulation of the bubble/bath system, and permits additional physical insights.
We minimize the energy using a trial wavefunction for the 1-2 bubble of the form:
\begin{equation}
\psi_{1,2}(r; \sigma,s)=A_{1,2}\exp\left[-\frac{1}{2}\left(\frac{r}{\sigma}\right)^{s}\right],
\end{equation} 
where the variational parameters $\sigma$ and $s$ characterize, respectively, the bubble radius, and the flatness of the bubble profile. The latter interpolates between 
a Gaussian ($s=2$) and a flat-top solution for $s\gg 2$~\cite{Lavoine2021}.  For the bath, we employ the variational form: 
\begin{equation}
\psi_{3}(r; r_{0},\delta r)=A_{3}\left[1+\tanh\left(\frac{r-r_{0}}{\delta r}\right)\right], 
\end{equation}
where the variational parameters $r_0$ and $\delta r$ characterize, respectively, the radius of the hollow cavity in the bath, and the bath healing length back into the homogeneous density value. The amplitudes $A_{1,2,3}$ are found upon normalization to the number of particles $N_{\sigma}=\int d^{3}r\left|\psi_{\sigma}(r)\right|^{2}$

\begin{figure}[t!]
\begin{centering}
\includegraphics[width=\columnwidth]{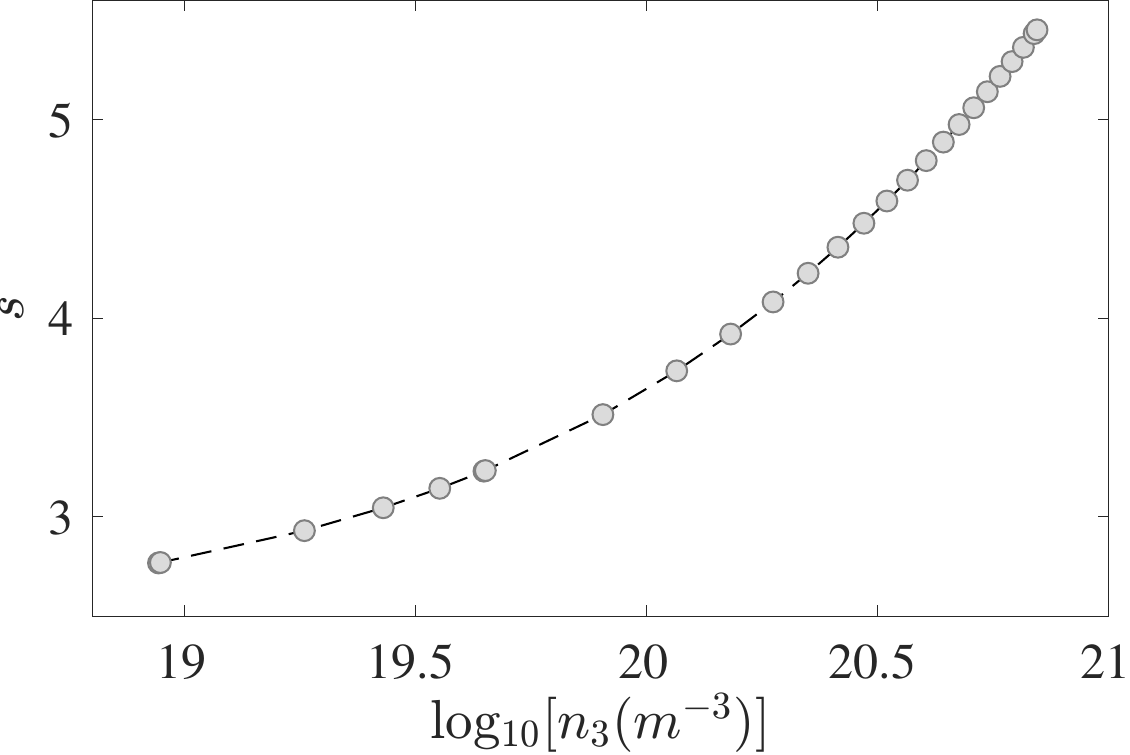}
\caption{Exponent $s$ that characterizes the profile of the droplet. $s=2$ is a Gaussian-like droplet, whereas $s\gg 2$ indicates a flat-top profile.
The calculations are performed for the same parameters as in Fig.~\ref{Fig:1}, with $N=10^{4}$ and $P=\sqrt{a_{11}/a_{22}}$. Note that when the bath density grows, the 
bubble profile becomes progressively more flat-top.}
\label{Fig:2}
\end{centering}
\end{figure}

\begin{figure}[t]
\begin{centering}
\includegraphics[width=\columnwidth]{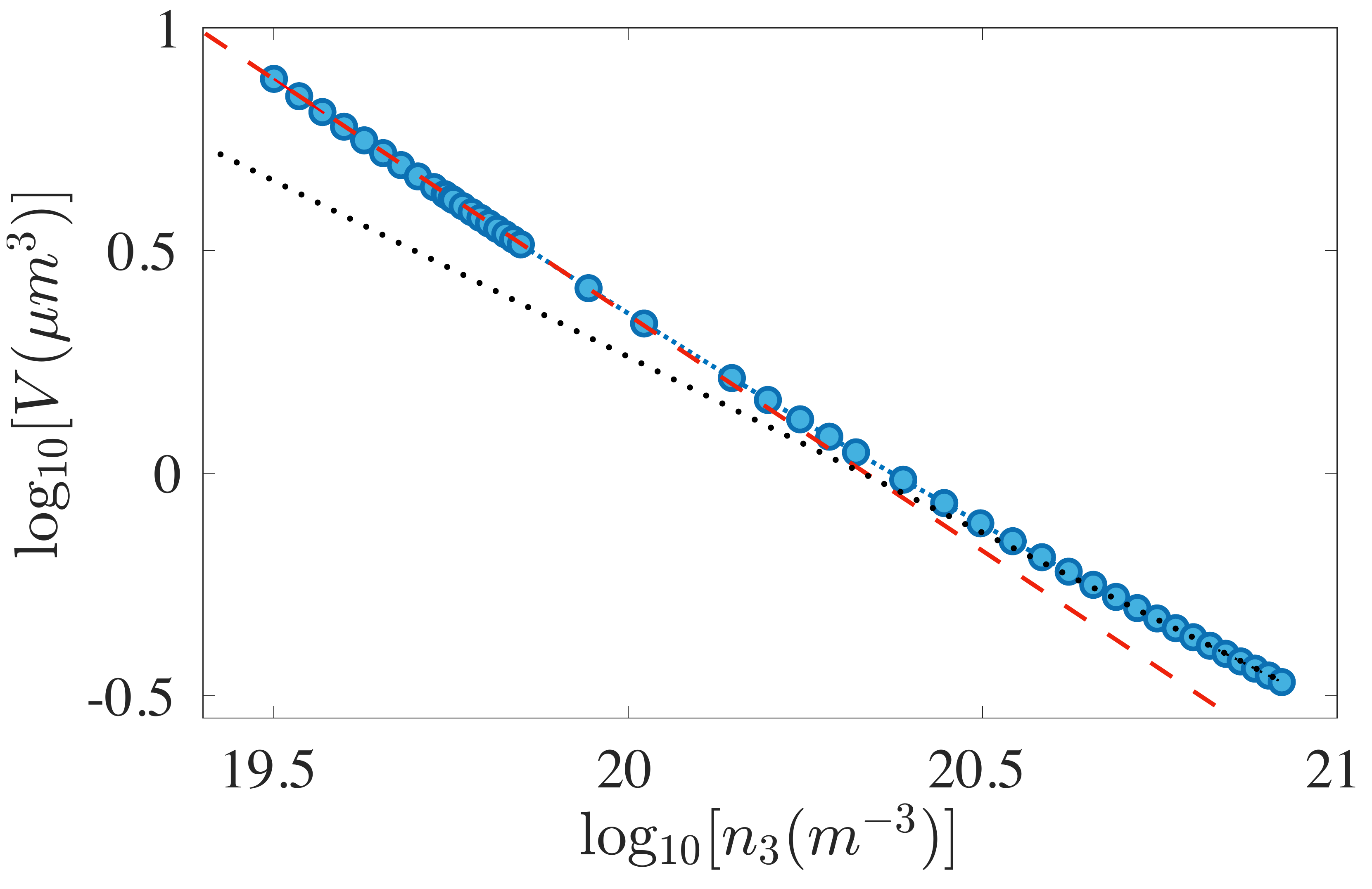}
\caption{Volume of the quantum bubble as a function of the bath density $n_3$ for the same scattering lengths and bubble polarization $P$ 
as in Fig.~\ref{Fig:1}, for $G(P)=0$ and $N=1650$. Note the crossover from a kinetic-energy dominated $V\propto n_3^{-6/5}$~(dashed line) dependence into a LHY-dominated 
$V\propto n_3^{-4/5}$ dependence~(dotted line).}
\label{Fig:3}
\end{centering}
\end{figure}

Figures~\ref{Fig:1}(a) and (b) depict our results for $G(P)=0$, $P=\sqrt{a_{11}/a_{22}}$,  $N=N_1+N_2=10000$, and two different bath densities, $n_3=9\times 10^{18}\mathrm{m}^{-3}$ and $7\times 10^{20}\mathrm{m}^{-3}$. We consider $(a_{11},a_{22},a_{13},a_{23},a_{33})=(34.44,82,172,172,60)a_{0}$~(see Sec.~\ref{Sec:Experiments}). Note that the variational calculations are in excellent qualitative and to a large extent quantitative agreement with the eGPE results. For a given number of particles $N$ in the bubble, increasing the bath density $n_3$, increases the outer pressure, compressing the bubble. In turn, the increase of the bubble density results in an enhanced role of interactions. Hence, for a LHY bubble ($G(P)=0$), when $n_3$ grows the bubble moves from a regime dominated by the kinetic energy~(as in Fig.~\ref{Fig:1}(a)) into a regime dominated by the LHY energy~(as in Fig.~\ref{Fig:1}(b)). As a result, the quantum bubble acquires a flat-top profile. The change in the character of the density profile is evident from Fig.~\ref{Fig:2}, where we plot the variational parameter $s$ as a function of the bath density for a fixed $N=10000$. Note the transition from a Gaussian-like profile $s\simeq 2$ to a flat-top, $s\gg 2$. 
According to the discussion above, the progressively larger role
played by the LHY energy for growing $n_3$ modifies the dependence of the bubble volume with the bath density. In Fig.~\ref{Fig:3}, we depict the volume as a function of the bath density, for $N=1650$ and the same scattering lengths as above. Note the expected crossover 
between a scaling $V\propto n_3^{-6/5}$ when the kinetic energy dominates, and a scaling $V\propto n_3^{-4/5}$, when the LHY energy dominates.



\section{Anomalous buoyancy}
\label{Sec:Buoyancy}

We investigated in the previous section a bubble in an otherwise homogeneous bath. At this point, we consider a more realistic situation, in which the mixture is 
confined in an isotropic harmonic trap, characterized by a frequency  $\omega$ for the 1-2 components, and a frequency $\omega_{3}$ for the bath. 
We assume that interactions in the bath are strong enough to result in a Thomas-Fermi radial density profile $n_{3}(r)=n_{3}(0)\left(1-r^{2}/R^{2}\right)$. 

The presence of a trap results eventually in buoyancy~\cite{Timmermans1998}. For a given ratio $\omega/\omega_3>\left ( \omega/\omega_3 \right )_{cr}$, the 1-2 bubble remains at the trap center. In contrast, when $\omega/\omega_3<\left ( \omega/\omega_3 \right )_{cr}$ the bubble moves from the center~\cite{Timmermans1998}. 
A mean-field bubble floats to the bath surface, where it is destroyed forming a partial or complete spherical shell around the bath. We show in this section that quantum fluctuations significantly modify the buoyancy condition. Moreover, they may lead to arrested buoyancy, i.e. the displacement of the bubble to an intermediate position 
between the center and the surface of the bath. Finally, for $\delta a<0$, when the bubble moves to the surface it does not spill over the surface, but undergoes a transition into a self-bound droplet that remains compact floating at the bath surface.

\subsection{Buoyancy condition for a uniform bubble}

We first consider the simplified case in which the bubble density is homogeneous within the hollow cavity, and in which the bubble volume is much smaller than the overall bath volume. Under these conditions, we may neglect the kinetic energy and the boundary effects associated to inter-particle interactions between 1-2 and 3 at the domain wall.
We can derive, as for the homogeneous case, the equation for the equilibrium of pressures for a droplet at position $r$:
\begin{equation}
G(P)n(r)^{2}+3\gamma(P) g_{11}a_{11}^{3/2}n(r)^{5/2}=g_{33}n_{3}(r)^{2}.
\label{eq:Press}
\end{equation}
The energy per particle of the bubble is:
\begin{eqnarray}
\frac{E(r)}{N}&=&\frac{1}{2}m\omega^{2}r^{2}+\frac{1}{2}G(P)n(r)
\nonumber \\
&+&\gamma (P)g_{11}a_{11}^{3/2}n(r)^{3/2}+\frac{1}{2}g_{33}\frac{n_{3}^{2}(r)}{n(r)}.
\label{eq:EN}
\end{eqnarray}
For $r\approx 0$, we can approximate $n(r) \simeq n(0) (1+\epsilon(r))$ with $\epsilon(r) \ll 1$. Plugging this expression into Eq.~\eqref{eq:EN}, using Eq.~\eqref{eq:Press} at $r=0$, 
as well as the Thomas-Fermi relation $\frac{1}{2}m\omega_3^{2}R^{2}=g_{33}n_{3}(0)$, we obtain
\begin{equation}
\frac{(E(r)-E(0))/N}{g_{33} n_3(0)} \simeq   \left [ \left ( \frac{\omega}{\omega_3} \right )^2-\frac{n_3(0)}{n(0)}\right ] \left ( \frac{r}{R} \right )^2.
\end{equation}
We hence obtain the critical frequency ratio for buoyancy:
\begin{equation}
 \left ( \frac{\omega}{\omega_3} \right )_{cr} =\sqrt{\frac{n_3(0)}{n(0)}}.
 \label{eq:ratio}
\end{equation}
We recall that for $\frac{\omega}{\omega_3}<\left ( \frac{\omega}{\omega_3} \right )_{cr}$ the bubble moves out of the trap center. However, as shown below, 
this does not necessarily mean that it moves to the bath surface.

\subsubsection{Mean-field bubble}

For a mean-field bubble, in which we can neglect the LHY contribution, the density ratio is given by Eq.~\eqref{eq:nMF}, and we retrieve the known critical frequency ratio for mean-field 
buoyancy~\cite{Timmermans1998}:
\begin{equation}
 \left ( \frac{\omega}{\omega_3} \right )_{cr}^{\mathrm{MF}} =\left ( \frac{G(P)}{g_{33}} \right )^{1/4}
\end{equation}

\subsubsection{LHY bubble}

For a LHY bubble ($G(P)=0$), the equilibrium of pressures leads to a simple dependence of $n(0)$ on $n_3(0)$ given by Eq.~\eqref{eq:nLHY}, and the critical 
 frequency ratio acquires the form 
\begin{equation} 
 \left ( \frac{\omega}{\omega_3} \right )_{cr}^{G=0}= \sqrt{\left(3\gamma(P)\right)^{2/5} \frac{a_{11}}{a_{33}}\left(n_{3}(0)a_{33}^{3}\right)^{1/5}}.
 \end{equation}
Note that it depends explicitly on the bath density. This is connected to the arrested buoyancy discussed below.

\subsubsection{Quantum bubble with $|\delta a|>0$}

The buoyancy condition may be obtained as well for the case of mean-field quasi-cancellation, when $\frac{|\delta a|}{\sqrt{a_{11}a_{22}}}\ll1$~(we assume $P=\sqrt{a_{11}/a_{22}}$). 
We can then expand in Eq.~\eqref{eq:f} $f(x,y=1/P)$ around  $x=1$, obtaining: 
\begin{equation}
 \gamma(P)\simeq \gamma_{0}(P)\left(1- \frac{5P}{(1+P)^{2}}\frac{\delta a}{\sqrt{a_{11}a_{22}}}\right)
\end{equation}
with $\gamma_{0}(P)=\frac{64}{15\sqrt{\pi}}P^{-5/2}$. At the critical frequency for buoyancy, we may substitute the relation~\eqref{eq:ratio}  in the 
equation for the equilibrium of pressures, obtaining:
\begin{equation}
    \frac{a(P)}{a_{33}}\left(\frac{\omega_{3}}{\omega}\right)_{cr}^{4}+3\gamma(P)\left(\frac{a_{11}}{a_{33}}\right)^{5/2}\sqrt{n_{3}(0)a_{33}^{3}}\left(\frac{\omega_{3}}{\omega}\right)_{cr}^{5}=1
    \label{eq:BUOY2}
\end{equation}
with $a(P)=\frac{2P}{(1+P)^{2}}\delta a$. We may then evaluate up to first order in $\frac{|\delta a|}{\sqrt{a_{11}a_{22}}}$ the critical frequency ratio for the buoyancy of a quantum bubble:
\begin{equation}
  \left(\frac{\omega}{\omega_{3}}\right)_{cr}^{G\neq 0}\simeq\left(\frac{\omega}{\omega_{3}}\right)_{cr}^{G=0}\left(1+\eta\right),
  \label{eq:frecuencyper}
\end{equation}
where 
\begin{equation}
   \eta= \frac{a(P)}{2\sqrt{a_{11}a_{22}}}\left[1-\frac{2}{5}\frac{\sqrt{a_{11}a_{22}}}{a_{33}} \left ( \left(\frac{\omega}{\omega_3}\right)_{cr}^{G=0}\right ) ^{-4}\right].
\end{equation}



\begin{figure}[t!]
\begin{centering}
\includegraphics[width=0.9\columnwidth]{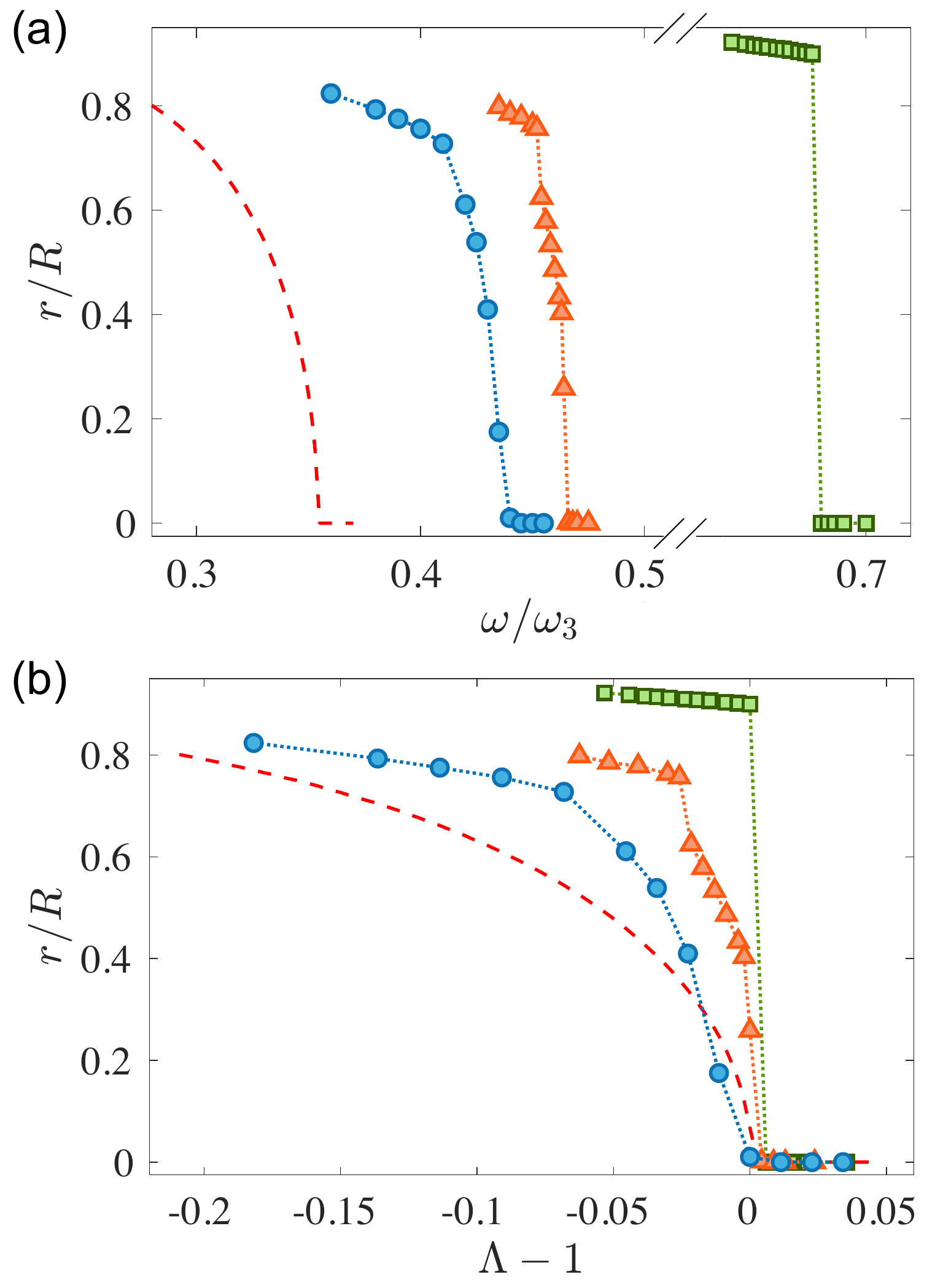}
\caption{(a) Position of the quantum bubble for a Thomas-Fermi bath with $n_3(0)=5\times10^{20}m^{-3}$ (using $N_3=8\times10^{5}$ in a trap with $\omega_3=2\pi \times 150$Hz), for the same parameters as Fig.~\ref{Fig:1}, for 
$N_1=1000$~(squares), $N_1=20000$~(triangles), $N_1=50000$~(circles) and $a_{33}=100a_0$. The dashed curve indicates the position of  the bubble obtained for the same $n_3(0)$ from the minimization of the energy per particle~\eqref{eq:EN}, using the relation~\eqref{eq:Press} between bubble and bath density. (b) Shows in detail the arrested buoyancy regime in the 
vicinity of the critical frequency ration for buoyancy, $\Lambda\equiv \frac{(\omega/\omega_3)}{(\omega/\omega_3)_{cr}}=1$. The arrested buoyancy window becomes more apparent 
when the LHY dominates the bubble physics. The scattering lengths considered are the same as in Fig.~\ref{Fig:1}, except that $a_{12}$ is slightly shifted, such that $\delta a=-5a_0$.}
\label{Fig:4}
\end{centering}
\end{figure}




\begin{figure*}[t]
\begin{centering}
\includegraphics[width=1.5\columnwidth]{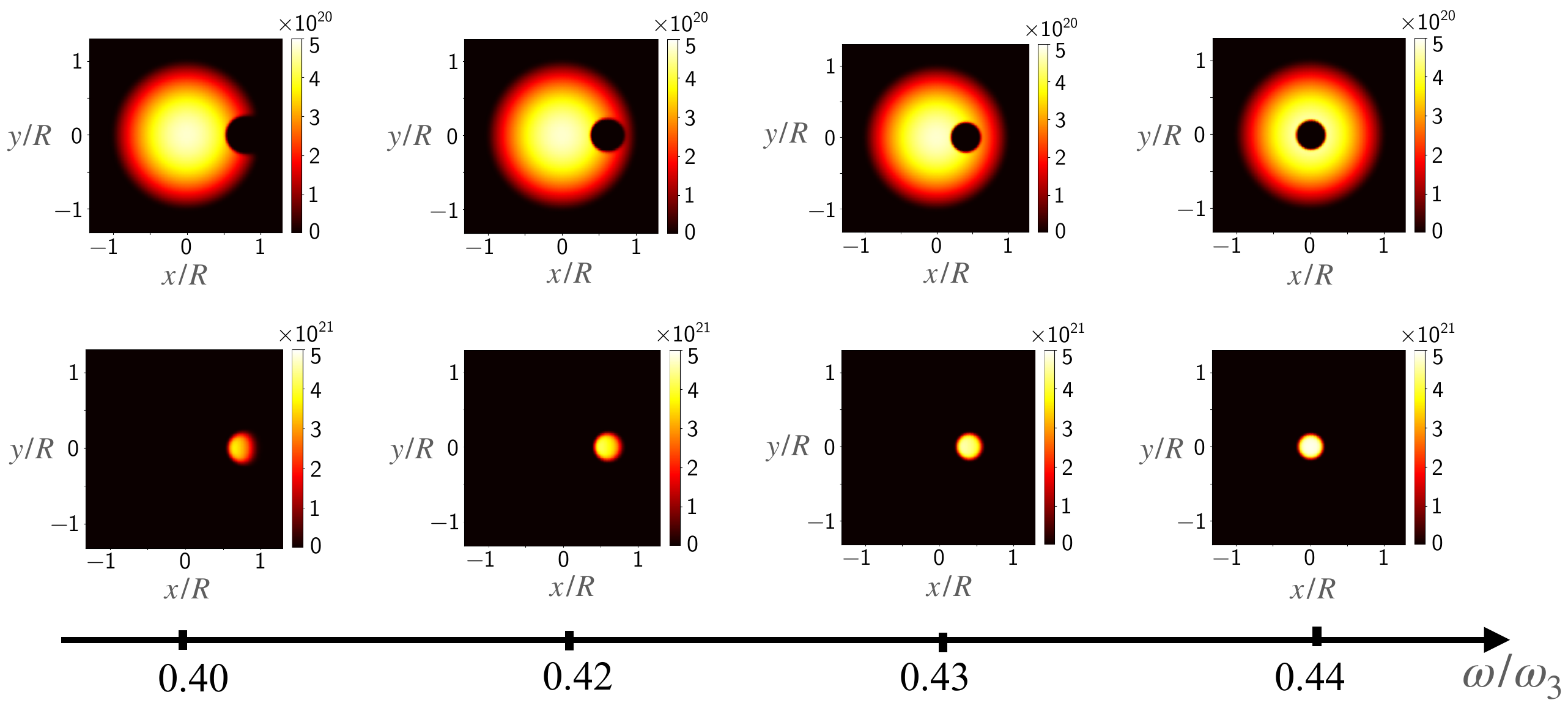}
\caption{Arrested buoyancy for the same parameters as Fig.~\ref{Fig:4}, for $N=8.2\times10^{4}$ and $n_3(0)=4.7\times10^{20}m^{-3}$. The~upper~(lower) panels show the density profile of the bath $n_3(x,y,0)$~(bubble $n_1(x,y,0)$). When buoyancy sets in, and due to the effect of quantum fluctuations, the bubble does not move immediately to the bath surface, but rather remains at an intermediate distance between the center and the surface, breaking spontaneously spherical symmetry. Note also that when the bubble moves to the surface it remains compact, experiencing a crossover into a self-bound droplet.}
\label{Fig:5}
\end{centering}
\end{figure*}


\subsection{Arrested buoyancy}

Interestingly, when occurring, buoyancy may differ significantly from the well-known mean-field case, due to the role played by quantum fluctuations in the 1-2 gas.
Let us first consider the case of a mean-field bubble.  Using the Thomas-Fermi form of $n_3(r)$, we re-write the bubble energy as:
\begin{equation}
\frac{E(r)/N}{g_{33} n_3(0)}= \left [\left ( \frac{\omega}{\omega_3} \right )_{cr}^{\textrm{MF}} \right ]^2 \left ( 1- \left (\frac{r}{R}\right )^2 \right )  + 
\left ( \frac{\omega}{\omega_3} \right )^2 \left ( \frac{r}{R} \right )^2.
\end{equation}
For $\omega/\omega_3 <  \left ( \omega/\omega_3 \right )_{cr}^{\textrm{MF}}$ the bubble passes from the center directly to the bath surface since the energy
becomes monotonously decreasing with increasing $r/R$. At the surface the mean-field bubble, which was solely maintained by the outer bath pressure, is destroyed and it forms a partial or complete covering of the bath spherical surface.

For an LHY bubble, we may express the bubble energy as:
\begin{equation}
\frac{E(r)/N}{g_{33} n_3(0)}=  \frac{5}{6} \left [ \left ( \frac{\omega}{\omega_3} \right )_{cr}^{G=0} \right ]^2 \left(1-\frac{r^{2}}{R^{2}}\right)^{6/5} + \left(\frac{\omega}{\omega_{3}}\right)^{2} \frac{r^{2}}{R^{2}}.
\label{eq:ENr}
\end{equation}

In contrast to the mean-field case, when $\Lambda \equiv \frac{\omega}{\omega_3} / \left ( \frac{\omega}{\omega_3} \right )_{cr}^{G=0} < 1$, the bubble energy has a minimum at 
\begin{equation}
\frac{r}{R} = \sqrt{1-\Lambda^{10} }
\end{equation}
Hence when buoyancy sets in, the position of the bubble does not immediately jump to the surface, as in the mean-field case, but rather 
experiences an abrupt, but finite, position displacement, breaking spontaneously the spherical symmetry. 
The red dashed curve in Fig.~\ref{Fig:4}(a) shows, as a function of $\frac{\omega}{\omega_3}$, the average position for an homogeneous quantum bubble, 
well within the quasi-cancellation regime for the parameters considered. As expected from the discussion above, there is a window of frequency ratios for which the 
bubble is placed at an intermediate position within the bath component. Note that in the regime of arrested buoyancy the bubble breaks the spherical symmetry of the model. 
We hence foresee an interesting superfluid rotation dynamics of quantum bubbles within the bath, which may occur at arbitrarily low energies. This discussion lies however beyond the scope of the present paper.

\subsection{Buoyancy for an inhomogeneous bubble}

The previous discussion neglects the kinetic energy of the bubble, which in general  may have a sizable contribution, and 
assumes that the bubble size is negligible with respect to the size of the Thomas-Fermi cloud of the bath. The latter is a particularly crude approximation 
under typical conditions. 

We have evaluated the ground-state of the mixture using the coupled eGPEs~\eqref{eq:Coupled-eGPE}, adding the confinement potential. 
The circles in Fig.~\ref{Fig:4}(a) shows our results of the 
average position of the bubble for a small number of particles $N_1=1000$. For this case, the LHY term is negligible and the bubble properties are dominated by the kinetic energy. 
As a result, compared to Eq.~\eqref{eq:frecuencyper}, a larger $\omega/\omega_3$ ratio is necessary to keep the bubble confined at the center. Also, when buoyancy sets in, there is no discernible regime of arrested buoyancy. 

When the number of particles in the bubble increases (or alternatively for a growing $n_3$) $(\omega/\omega_3)_{cr}$ decreases, and a 
progressively wider window of arrested buoyancy is observed~(see Fig.~\ref{Fig:4}(b)). Although the numerical results approach the result of the homogeneous-droplet calculation, there are still 
sizable deviations of the critical frequency ratio compared to Eq.~\eqref{eq:frecuencyper}, mostly due to the non-negligible size of the bubble compared to the Thomas-Fermi cloud of the bath. 

Figure~\ref{Fig:5} shows the density profile of the mixture in the arrested buoyancy regime for $N_1=50000$ atoms. Note that, as discussed for the case of an homogeneous droplet, the inhomogeneous bubble is placed at intermediate positions (spontaneously breaking the spherical symmetry). Note as well, that in contrast to the mean-field case, when the bubble reaches the boundary, it does not spread around the spherical Thomas-Fermi surface. Since $\delta a <0$, it rather undergoes a crossover from a bubble into a self-bound droplet, which remains compact floating at the bath surface.




\section{Experimental considerations}
\label{Sec:Experiments}

A possible implementation of the quantum bubble scenario discussed in this paper is provided by the multi-component $^{41}$K $-$ $^{39}$K mixture~\cite{ThesisCRCC}, whose 
concrete parameters have been employed in our simulations. In this implementation, 
the 1-2 gas is composed by a $^{39}$K mixture in states $\ket{1} \equiv \ket{F=1, m_F =-1}$ and  
$\ket{2} \equiv \ket{F=1, m_F =0}$, whereas the bath is composed by the state $\ket{3} \equiv  \ket{F=1, m_F =-1}$ of $^{41}$K. In this setting, the system is in the lowest energy state and inelastic spin-exchange collisions can be neglected~\cite{LTanzi}. For the bubble, the parameter $\delta a \leq$ 0 can be tuned in the vicinity of $\sim 56.9$ G~\cite{Cabrera2018, Semeghini2018} where the overlap of three different Feshbach resonances allows to control the values of $a_{11}$, $a_{22}$, and $a_{12}$. At  this magnetic field, the bath-bubble interactions is set by the background $^{41}$K-$^{39}$K scattering length, which is constant ($a_{13}=a_{23}\approx 172 a_0$). 

Since the system is composed by two different potassium isotopes, high-resolution \textit{in situ} imaging detection can be performed in order to extract the bubble and bath density profiles independently. Experiments may then readily monitor how the contribution of the LHY energy at $G=0$ affects the bubble size. For typical densities of $n_3 =10^{20} \mathrm{m}^{-3}$, we can expect for large atom number in the bubble a discrepancy of up to $40\%$ in its radius compared to the case where quantum fluctuations are neglected. This discrepancy becomes larger when increasing the density $n_3$. Hence, the analysis of the bubble size may readily reveal the effect of quantum fluctuations and the scaling features 
discussed in this paper.




\section{Conclusions}
\label{Sec:Conclusions}

We have considered a peculiar effective immiscible binary mixture. Two miscible components form an effective scalar condensate (1-2 gas) 
with enhanced quantum fluctuations due to mean-field quasi-cancellation, and a third component is immiscible with the other two. We have shown that due to quantum fluctuations, the properties of the effective mixture significantly depart from those well-known for an immiscible mean-field Bose-Bose gas.
In particular, the volume of a quantum 1-2 bubble in component 3 presents a significantly modified dependence with respect to the bath density. Moreover, quantum fluctuations lead to an anomalous buoyancy criterion. Once buoyancy sets in, in contrast to the case of mean-field mixtures, the bubble may occupy an intermediate position between the center and the surface of the bath~(arrested buoyancy). Furthermore, once the surface is reached the bubble may transition into a droplet, which remains compact and floating at the bath surface. 

These results, which may be readily probed in e.g. Potassium mixtures, illustrate how quantum fluctuations, in addition to providing the stabilization mechanism for self-bound droplets, may significantly change other general and well established properties of Bose mixtures. We anticipate that a similar physics may be at play as well in immiscible mixtures, in which at least one of the components (forming the quantum bubble) is dipolar within the regime of mean-field quasi-cancellation~\cite{Bland2022}. However, the nonlocal anisotropic character of the dipole-dipole interaction may significantly affect the droplet properties and the buoyancy condition. \\




\acknowledgements
We acknowledge support of the Deutsche Forschungsgemeinschaft (DFG, German Research Foundation) under Germany’s Excellence Strategy– EXC-2123 QuantumFrontiers– 390837967, and FOR 2247. C.R.C has received funding from the European Union’s Horizon 2020 research and innovation programme under the Marie Skłodowska-Curie grant agreement No 897142.

\end{document}